\documentclass[a4paper,onecolumn,11pt,accepted=2023-04-06]{quantumarticle}
\pdfoutput=1
\usepackage[utf8]{inputenc}
\usepackage[english]{babel}
\usepackage[T1]{fontenc}
\usepackage{amsmath}
\usepackage{hyperref}

\usepackage{tikz}
\usepackage{lipsum}

\begin{document}

\title{Signature of exceptional point phase transition in Hermitian systems}

\author{Timofey T. Sergeev}
\affiliation{Dukhov Research Institute of Automatics, 127055,  22 Sushchevskaya, Moscow Russia}
\affiliation{Moscow Institute of Physics and Technology, 141700, 9 Institutskiy pereulok, Dolgoprudny Moscow region, Russia}
\affiliation{Institute for Theoretical and Applied Electromagnetics, 125412, 13 Izhorskaya, Moscow Russia}

\author{Alexander A. Zyablovsky}
\email{zyablovskiy@mail.ru}
\affiliation{Dukhov Research Institute of Automatics, 127055, 22 Sushchevskaya, Moscow Russia}
\affiliation{Moscow Institute of Physics and Technology, 141700, 9 Institutskiy pereulok, Dolgoprudny Moscow region, Russia}
\affiliation{Institute for Theoretical and Applied Electromagnetics, 125412, 13 Izhorskaya, Moscow Russia}
\affiliation{Kotelnikov Institute of Radioengineering and Electronics RAS, 125009, 11-7 Mokhovaya, Moscow Russia}

\author{Evgeny S. Andrianov}
\affiliation{Dukhov Research Institute of Automatics, 127055, 22 Sushchevskaya, Moscow Russia}
\affiliation{Moscow Institute of Physics and Technology, 141700, 9 Institutskiy pereulok, Dolgoprudny Moscow region, Russia}
\affiliation{Institute for Theoretical and Applied Electromagnetics, 125412, 13 Izhorskaya, Moscow Russia}

\author{Yurii E. Lozovik}
\affiliation{Institute of Spectroscopy Russian Academy of Sciences, 108840, 5 Fizicheskaya, Troitsk, Moscow, Russia}
\affiliation{MIEM at National Research University Higher School of Economics, 101000, 20 Myasnitskaya, Moscow, Russia}

\maketitle

\begin{abstract}
  Exceptional point (EP) is a spectral singularity in non-Hermitian systems. The passing over the EP leads to a phase transition, which endows the system with unconventional features that find a wide range of applications. However, the need of using the dissipation and amplification limits the possible applications of systems with the EP. In this work, we demonstrate an existence of signature of exceptional point phase transition in Hermitian systems that are free from dissipation and amplification. We consider a composite Hermitian system including both two coupled oscillators and their environment consisting only of several tens of degrees of freedom. We show that the dynamics of such a Hermitian system demonstrate a transition, which occurs at the coupling strength between oscillators corresponding to the EP in the non-Hermitian system. This transition manifests itself even in the non-Markovian regime of the system dynamics in which collapses and revivals of the energy occur. Thus, we demonstrate that the phase transition occurring at the passing over the EP in the non-Hermitian system manifests itself in the Hermitian system at all time. We discuss the experimental scheme to observe the signature of EP phase transition in the non-Markovian regime.
\end{abstract}

Non-Hermitian systems are actively investigated in last decades \cite{ref1}-\cite{ref2-2}. Unlike the Hermitian systems, the eigenstates of non-Hermitian systems are no mutually orthogonal \cite{ref1,ref2}. The point in the space of system parameters, at which some of eigenstates coalesce and corresponding eigenvalues coincide, is called an exceptional point (EP) of non-Hermitian system \cite{ref1,ref2}. By changing parameters, the system can pass over the EP. This passing is accompanied by qualitative change in the eigenstates \cite{ref3,ref4}, which is referred to as an exceptional point phase transition \cite{ref2},\cite{ref5}-\cite{ref10}. An example of such a transition is the spontaneous symmetry breaking in PT-symmetry systems \cite{ref4},\cite{ref11}-\cite{ref13}. The EP phase transitions also take place in strongly coupled cavity-atom \cite{ref3,ref14,ref20}, polaritonic \cite{ref15}-\cite{ref17}, optomechanical \cite{ref18,ref19}, and parametrically driven \cite{ref19-2,ref37} systems. In the vicinity of the phase transition point the non-Hermitian systems demonstrate unusual properties, which find a numerous applications \cite{ref3,ref4,ref12,ref13}. In particular, they are used to enhance the sensitivity of laser gyroscopes \cite{ref21} and sensors \cite{ref22}-\cite{ref25}, to control light propagation \cite{ref26}-\cite{ref32}, to control the laser characteristics \cite{ref33}-\cite{ref36} and even to achieve lasing without population inversion \cite{ref37,ref38}.

In spite of the significant advances in the applications of the systems with the EP, the necessity of using the gain and loss limits the utility of these devices \cite{ref39}-\cite{ref46}. For instance, the dissipation as well as amplification leads to increase of noises near the EP that prevents to sensor operation \cite{ref39}-\cite{ref44-2}. Moreover, the dissipation (amplification) impedes a stable device operation due to the decay (growth) of energy in the systems \cite{ref45,ref46}. Therefore, overcoming the negative influence of dissipation and amplification on the characteristics of the devices with the EP is an important problem.

In this paper, we consider a Hermitian system consisting of a small subsystem and its environment. The interaction of the subsystem with environment leads to energy exchange between them. In the case of environment with infinite number of degrees of freedom, the energy irreversibly flows from the subsystem to the environment, which leads to energy dissipation in the subsystem. In the case of environment with large but finite number of degrees of freedom, situation qualitatively changes. At small times, an exponential decay of the amplitude oscillations in this subsystem occurs, which is due to the energy flows from the subsystem into the environment. However, at large times, the inverse energy flow from the environment leads to the revivals of oscillations in the subsystem \cite{ref47,ref48}. The time of the first appearance of the revival can be called as a return time \cite{ref49}. Usually, it is assumed that the environment is so large that the return time is much longer than observation time, and revivals can be excluded from consideration. In this case, the subsystem interacting with the environment can be described by non-Hermitian equations. However, despite the fact that the subsystem demonstrates the non-Hermitian dynamics at times smaller than the return time, the eigenstates of whole Hermitian system including the environment are always mutually orthogonal. Since the eigenstates of Hermitian systems are mutually orthogonal \cite{ref2}, the EPs cannot take place in such systems. The phase transitions causing the transition through the EP are many times observed in the different non-Hermitian systems \cite{ref3,ref4}. However, their existence in Hermitian system early is not discussed so far.

Here, we demonstrate that the signature of EP phase transition can be observed at time much greater the return time. As an example, we consider a Hermitian system including both two coupled oscillators and the environment consisting only of several tens of degrees of freedom. We demonstrate that the dynamics of such a Hermitian system exhibit a transition at the change in the coupling strength between the oscillators. This transition manifests itself even at times much greater than the return time, when the system displays the non-Markovian dynamics including revivals of the energy of the oscillators. We show that there is a threshold coupling strength between the oscillators. Below the threshold coupling strength, the system evolves from any initial state into a state with a strictly specified ratio of the oscillators’ amplitudes. Above the threshold coupling strength, the ratio of the oscillators’ amplitudes in the final state depends on the initial state. To illustrate this fact, we calculate the dependence of the variance of the amplitudes’ ratio by varying the initial states on the coupling strength. We demonstrate that this variance is about zero below the threshold coupling strength and sharply increases when the coupling strength exceeds the threshold value. Therefore, the variance of the amplitudes’ ratio can serve as an order parameter of the transition. We show that the threshold coupling strength in the Hermitian system coincides with the one corresponding to the EP in the non-Hermitian system. We propose a photonic circuit based on two coupled microcavities interacting with finite-length waveguide for observation of the signature of EP phase transition in non-Markovian regime.

The obtained results can help to overcome the limitations arising from amplification and dissipation for the practical utilization of systems with the EP phase transitions. This opens the way for creation of a new type of devices for laser and sensoric applications.

\section{System under consideration}
\begin{figure}[t]
  \centering
 \includegraphics[width=0.7\linewidth]{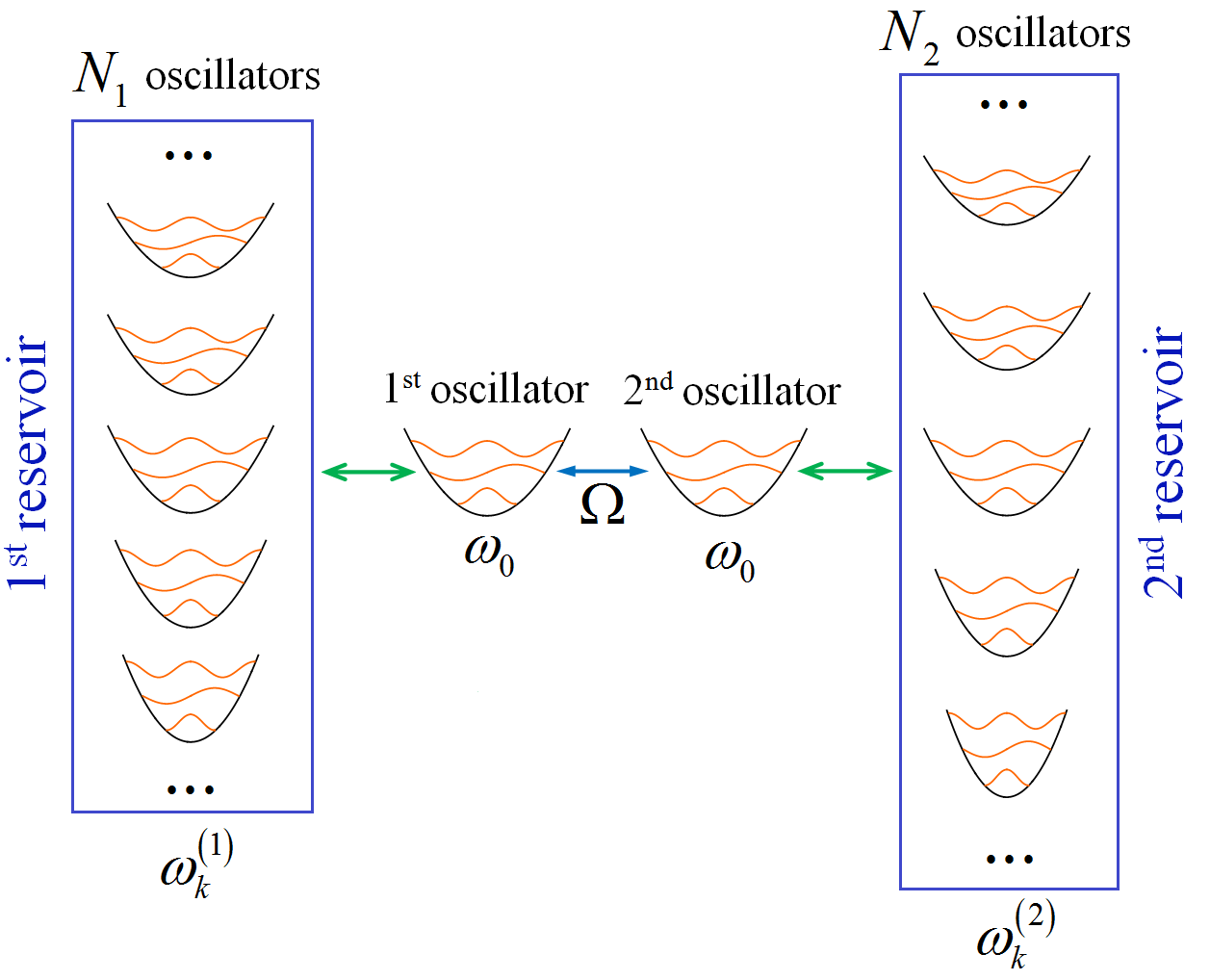}
  \caption{A scheme of Hermitian system under consideration consisting of two coupled oscillators and two reservoirs of  ${N_{1,2}}$ oscillators. Each of the oscillators interacts with its reservoir.}
  \label{fig:figure1}
\end{figure}

We consider a Hermitian system which includes two coupled oscillators and two reservoirs, each of which interacts with one of oscillators (Figure~\ref{fig:figure1}). The frequencies of two coupled oscillators are equal to ${\omega _0}$. The first and second reservoirs consist of sets of ${N_{1,2}}$ oscillators with frequencies $\omega _k^{\left( 1 \right),\left( 2 \right)} = {\omega _0} + \delta {\omega _{1,2}}\,(k - {N_{1,2}}/2)$, respectively. Here $\delta {\omega _1}$ and $\delta {\omega _2}$ are steps between the oscillator frequencies in the first and second reservoirs.

To describe this system, we use the Hamiltonian of interacting oscillators in the rotating-wave approximation \cite{ref49}:

\begin{equation}
\label{eq1}
 \begin{array}{l}
\hat H = {\omega _0}\hat a_1^\dag {{\hat a}_1} + {\omega _0}\hat a_2^\dag {{\hat a}_2} + \Omega (\hat a_1^\dag {{\hat a}_2} + \hat a_2^\dag {{\hat a}_1}) + \sum\limits_{k = 1}^{{N_1}} {\omega _k^{\left( 1 \right)}\hat b_k^\dag {{\hat b}_k}}  + \sum\limits_{k = 1}^{{N_2}} {\omega _k^{\left( 2 \right)}\hat c_k^\dag {{\hat c}_k}} \\
 + \sum\limits_{k = 1}^{{N_1}} {{g_1}(\hat b_k^\dag {{\hat a}_1} + \hat a_1^\dag {{\hat b}_k})}  + \sum\limits_{k = 1}^{{N_2}} {{g_2}(\hat c_k^\dag {{\hat a}_2} + \hat a_2^\dag {{\hat c}_k})} 
\end{array}
\end{equation}
First, second and third terms are the Hamiltonians of the first and second oscillators and the interaction between them, respectively. ${\hat a_{1,2}}$ and $\hat a_{1,2}^\dag$ are the annihilation and creation operators for the first and second oscillators, obeying the boson commutation relations \cite{ref49}. $\Omega$ is a coupling strength between the oscillators. Fourth and fifth terms are the Hamiltonians of the reservoirs interacting with the first and second oscillators, respectively. ${\hat b_k}$, ${\hat c_k}$ and $\hat b_k^\dag $, $\hat c_k^\dag $ are the annihilation and creation operators for the oscillators with the frequency $\omega _k^{\left( 1 \right),\left( 2 \right)}$ in the first and second reservoirs, respectively. $\sum\limits_{k = 1}^{{N_1}} {{g_1}(\hat b_k^\dag {{\hat a}_1} + \hat a_1^\dag {{\hat b}_k})}$ and $\sum\limits_{k = 1}^{{N_2}} {{g_2}(\hat c_k^\dag {{\hat a}_2} + \hat a_2^\dag {{\hat c}_k})}$ are the Hamiltonians of the interaction between the oscillators and their reservoirs. ${g_{1,2}}$ are the coupling strengths between the respective oscillator and the oscillators in the reservoirs. Such a model describes a wide class of systems interacting with reservoirs \cite{ref49,ref50}. We propose a photonic realization of this system in Section 3.

Using the Heisenberg equation for operators \cite{ref50,ref51}, we obtain the closed system of equations for operators ${\hat a_{1,2}}$, ${\hat b_k}$, ${\hat c_k}$. Moving from the operators to their averages, we obtain the linear system of equations

\begin{equation}
\label{eq2}
\frac{{d{a_1}}}{{dt}} =  - i{\omega _0}{a_1} - i\,\Omega {a_2} - i\sum\limits_{k = 1}^{{N_1}} {{g_1}{b_k}}
\end{equation}

\begin{equation}
\label{eq3}
\frac{{d{a_2}}}{{dt}} =  - i{\omega _0}{a_2} - i\,\Omega {a_1} - i\sum\limits_{k = 1}^{{N_2}} {{g_2}{c_k}}
\end{equation}

\begin{equation}
\label{eq4}
\frac{{d{b_k}}}{{dt}} =  - i\omega _k^{\left( 1 \right)}{b_k} - i\,{g_1}{a_1}
\end{equation}

\begin{equation}
\label{eq5}
\frac{{d{c_k}}}{{dt}} =  - i\omega _k^{\left( 2 \right)}{c_k}\, - i\,{g_2}{a_2}
\end{equation}
where ${a_1} = \left\langle {{{\hat a}_1}} \right\rangle$, ${a_2} = \left\langle {{{\hat a}_2}} \right\rangle$, ${b_k} = \left\langle {{{\hat b}_k}} \right\rangle$ and ${c_k} = \left\langle {{{\hat c}_k}} \right\rangle$. Underline that Eqns.~\eqref{eq2}–\eqref{eq5} are a closed system for the operator averages and thus can be solved without additional approximations.

\section{Results}
\subsection{Crossover between Hermitian and non-Hermitian systems}
\begin{figure}[t]
  \centering
 \includegraphics[width=\linewidth]{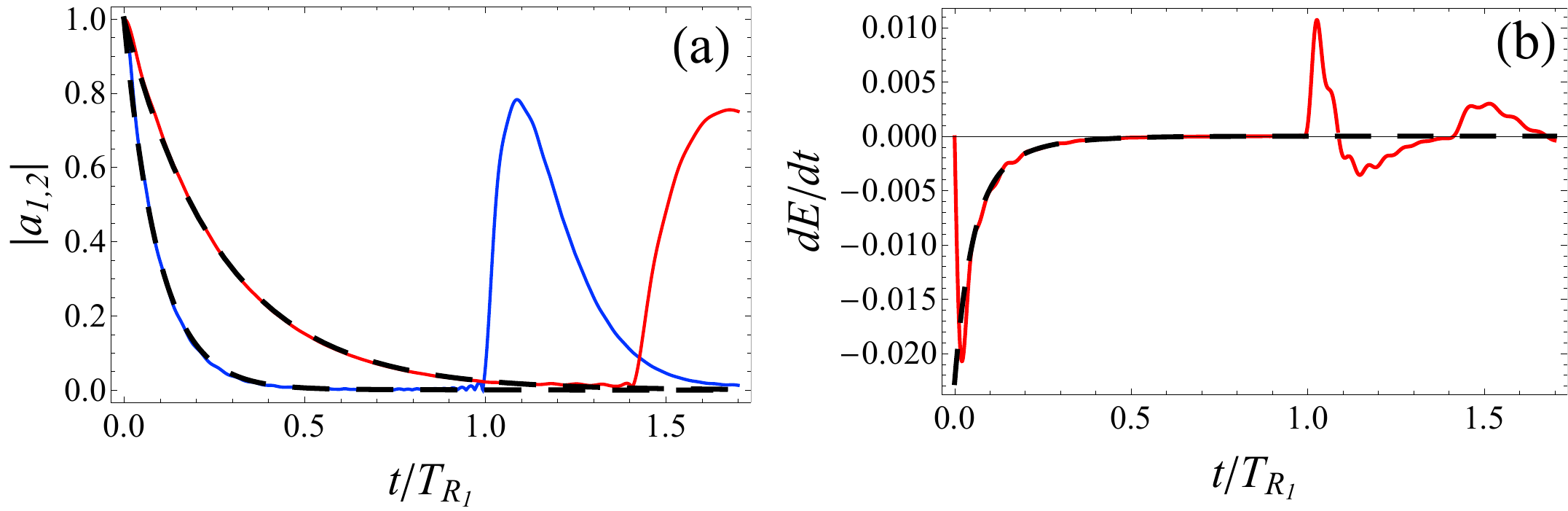}
  \caption{(a) Dependencies of the absolute values of the amplitudes of the first (blue line) and second (red line) oscillators on time calculated by simulating Eqns.~\eqref{eq2}–\eqref{eq5}. (b) Dependence of the time derivative energy of the subsystem of the two coupled oscillators (red line) on time calculated by simulating Eqns.~\eqref{eq2}–\eqref{eq5}. The energy of the subsystem is calculated as $E=\omega_0 {\left| {a_1} \right|^2} + \omega_0 {\left| {a_2} \right|^2} + \Omega (a_1^* a_2 + a_1 a_2^*)$. The black dashed lines show temporal dependencies calculated by Eq.~\eqref{eq6} with decay rates~\eqref{eq10}. $\Omega  < {\Omega _{EP}} = \frac{{{\gamma _1} - {\gamma _2}}}{2}$; ${T_{R1}} = \frac{{2\pi }}{{\delta {\omega _1}}}$ is the return time of the first reservoir; the return time of the second reservoir ${T_{R2}} = \sqrt 2 \,{T_{R1}}$. Here the following parameters are used: ${N_1} = {N_2} = 40$; $\delta {\omega _1} = 5 \times {10^{ - 3}}\,{\omega _0}$; $\delta {\omega _2} = 5 \times {10^{ - 3}}\,{\omega _0}/\sqrt 2$; $\Omega  = {10^{ - 4}}{\omega _0}$; ${g_{1}} \approx 3.6 \times {10^{ - 3}}\,{\omega _0}$; ${g_{1}} \approx 1.8 \times {10^{ - 3}}\,{\omega _0}$. The initial conditions are ${a_1}\left( t=0 \right) = 1$; ${a_2}\left( t=0 \right) = 1$; ${b_k}\left( t=0 \right) = 0$; ${c_k}\left( t=0 \right) = 0$.}
  \label{fig:figure2}
\end{figure}

The interaction of the oscillators with their reservoirs results in the energy exchange between them. When the number of degrees of freedom in the reservoirs tends to infinity (${N_{1,2}} \to \infty$ and $\delta {\omega _{1,2}} \to 0$), the reservoirs degrees of freedom can be eliminated from the consideration within the Born-Markovian approximation \cite{ref49,ref50,ref51}. In this approximation, it is considered that the reservoir is much larger than the system that the influence of the system on the states of the reservoir can be neglected \cite{ref50,ref51}. It is also assumed that the reservoir is in thermal equilibrium with temperature $T$ (for an empty reservoir i.e. when ${b_k}\left( t=0 \right) = 0$; ${c_k}\left( t=0 \right) = 0$, $T=0$). The flow of energy from the system to the reservoir depends on the state of the system, and the reverse flow is determined by the temperature of the reservoir \cite{ref50,ref51}. At zero temperature, the reverse flow is zero at all times \cite{ref50,ref51}. In this case,  the interaction of the oscillators with the reservoirs leads to an exponential decay of the oscillators amplitudes (Figure~\ref{fig:figure2}a). 

In the Born-Markovian approximation and $T=0$, the system dynamics are described by the effective non-Hermitian equation \cite{ref49,ref52}:

\begin{equation}
\label{eq6}
\frac{d}{{dt}}\left( {\begin{array}{*{20}{c}}
{{a_1}}\\
{{a_2}}
\end{array}} \right) = \left( {\begin{array}{*{20}{c}}
{ - i{\omega _0} - {\gamma _1}}&{ - i\Omega }\\
{ - i\Omega }&{ - i{\omega _0} - {\gamma _2}}
\end{array}} \right)\left( {\begin{array}{*{20}{c}}
{{a_1}}\\
{{a_2}}
\end{array}} \right)
\end{equation}
where ${\gamma _{1,2}}$ are effective decay rates, which, in the case of infinite reservoirs, are given as ${\gamma _{1,2}} = \sum\limits_{k = 1}^{{N_{1,2}}} {\pi g_{1,2}^2} \delta \left( {{\omega _0} - \omega _k^{\left( 1 \right),\left( 2 \right)}} \right)$, where $\delta \left( {\omega} \right)$ is Dirac’s delta-function \cite{ref49}. This system has an EP, at which the eigenstates coalesce and their eigenvalues coincide with each other. It occurs at

\begin{equation}
\label{eq7}
\Omega  =  \pm {\Omega _{EP}} =  \pm \frac{{\left| {{\gamma _1} - {\gamma _2}} \right|}}{2}
\end{equation}
The passing through the EP is accompanied by a spontaneous symmetry breaking of eigenstates,

\begin{equation}
\label{eq8}
{{\bf{h}}_{1,2}} = {\left( {\begin{array}{*{20}{c}}
{i\left( {\frac{{{\gamma _1} - {\gamma _2}}}{2} \pm \sqrt {\frac{{{{\left( {{\gamma _1} - {\gamma _2}} \right)}^2}}}{2} - {\Omega ^2}} } \right),}&\Omega 
\end{array}} \right)^T}
\end{equation}
and by a change of the eigenvalues ${\lambda _{1,2}} = - i {\omega_0} - \frac{{{\gamma _1} + {\gamma _2}}}{2} \pm \sqrt {\frac{{{{\left( {{\gamma _1} - {\gamma _2}} \right)}^2}}}{2} - {\Omega ^2}}$.
Below the EP ($\left| \Omega  \right| < {\Omega _{EP}}$), the eigenstates are non-PT-symmetrical and have different decay rates (${\mathop{\rm Re}\nolimits} {\lambda _1} \ne {\mathop{\rm Re}\nolimits} {\lambda _2}$). Above the EP ($\left| \Omega  \right| \ge {\Omega _{EP}}$), the eigenstates are PT-symmetrical and their decay rates become equal to each other (${\mathop{\rm Re}\nolimits} {\lambda _1} = {\mathop{\rm Re}\nolimits} {\lambda _2}$). The spontaneous symmetry breaking occurring at the EP is referred to as an EP phase transition \cite{ref2},\cite{ref5}-\cite{ref10}. The EP phase transition becomes apparent in the system dynamics \cite{ref1,ref6,ref53}, which are determined by as follows

\begin{equation}
\label{eq9}
\left( {\begin{array}{*{20}{c}}
{{a_1}}\\
{{a_2}}
\end{array}} \right) = {c_1}{{\bf{h}}_1}{e^{{\lambda _1}t}} + {c_2}{{\bf{h}}_2}{e^{{\lambda _2}t}}
\end{equation}
where ${c_1}$ and ${c_2}$ are amplitudes of eigenstates which are determined from the initial conditions. 
The dynamics are sensitive to the changes in the eigenstates and the eigenvalues. Often, it is the system dynamics that are the object of study when detecting the EP phase transition \cite{ref29,ref31,ref32}.

At finite sizes of reservoirs, i.e., when ${N_{1,2}}$ are finite and $\delta {\omega _{1,2}}/{\omega _0}$ are small but nonzero, from Eqns.~\eqref{eq2}–\eqref{eq5} it follows that the exponential decay takes place only when time is much smaller than the return times of both reservoirs, ${T_{R1,R2}} = 2\pi /\delta {\omega _{1,2}}$ \cite{ref67}. At return time ${T_{R1}}$ (or ${T_{R2}}$), the reservoirs' oscillators, $b_k$ (or $c_k$), constructively interfere with the same phases as at the initial moment of time and excite the oscillators' amplitudes, $a_{1,2}$. This occurs for the first time when the phase shift between the neighboring oscillators is $2\pi $. In the case of the reservoirs' oscillators with equidistant frequencies, $\omega _k^{\left( 1 \right),\left( 2 \right)} = {\omega _0} + \delta {\omega _{1,2}}\left( {k - {N_{1,2}}} \right)$, the phase difference between neighboring oscillators is $2\pi $, when $t=2 \pi/\delta {\omega _{1,2}}$ (${\omega _k^{\left( 1 \right),\left( 2 \right)} - {\omega _{k-1}}^{\left( 1 \right),\left( 2 \right)}} = \delta {\omega _{1,2}}$). Therefore, the return times are ${T_{R1,R2}} = 2\pi /\delta {\omega _{1,2}}$ \cite{ref67}.

At $t <  < {T_{R1,R2}}$ the temporal behavior of the oscillators interacting with the reservoirs of the finite number of degrees of freedom coincides with the behavior predicted by the non-Hermitian Eq.~\eqref{eq6} (cf. the blue and red lines with the dashed black line in Figure~\ref{fig:figure2}a) with the decay rates, which can be evaluated as

\begin{equation}
\label{eq10}
{\gamma _{1,2}} = \frac{{\pi g_{1,2}^2}}{{\delta {\omega _{1,2}}}}
\end{equation}

At $t > {T_{R1,R2}}$, the temporal dynamics of the oscillators interacting become more complex (Figure~\ref{fig:figure2}a). The revivals of the oscillations in the system occur at times $t \sim n\,{T_{R1,R2}}$, where $n$ is a natural number \cite{ref67}. When revival occurs, the energy flows from the reservoirs to the system (Figure~\ref{fig:figure2}b). This behavior does not take place in the non-Hermitian systems describing by Eq.~\eqref{eq6} (cf. the blue and red lines with the dashed black line in Figure~\ref{fig:figure2}b).

Thus, the dynamics of the two coupled oscillators interacting with reservoirs of finite sizes can be described by the non-Hermitian Eq.~\eqref{eq6} only at times much smaller than the return time. For this reason, it is expected that an analog of the EP phase transition can be manifested in the behavior of Hermitian system only at small times. However, the question arises "is the EP phase transition manifested at time much greater than the return time?" Below, we show that the signature of EP phase transition manifests itself in the system dynamics even at $t >> {T_{R1,R2}}$. We provide a new approach to describe the EP phase transition and introduce an order parameter that is suitable for both the non-Hermitian and Hermitian systems. Then, we demonstrate an existence of the signature of EP phase transition in the dynamics of entirely Hermitian system including the reservoirs with only several tens degrees of freedom.

\subsection{Criterion of phase transition at the EP}
To proceed, we consider dynamics of the non-Hermitian system~\eqref{eq6} both below and above the EP. The EP phase transition is accompanied by the change in the relaxation rates of the eigenstates. Below the EP, the eigenstates of the non-Hermitian system~\eqref{eq6} have different decay rates (to be specific, we consider ${\mathop{\rm Re}\nolimits} {\lambda _1} > {\mathop{\rm Re}\nolimits} {\lambda _2}$). In this case, at $t >  > {\left| {{\mathop{\rm Re}\nolimits} {\lambda _1} - {\mathop{\rm Re}\nolimits} {\lambda _2}} \right|^{ - 1}}$ the contribution of the first eigenstate to the system state becomes prevailing (see Eq.~\eqref{eq9}). As a result, at $t >  > {\left| {{\mathop{\rm Re}\nolimits} {\lambda _1} - {\mathop{\rm Re}\nolimits} {\lambda _2}} \right|^{ - 1}}$ the ratio of amplitudes of the first and the second oscillators, ${a_1}/{a_2}$, coincides with the one for the first eigenstate (see Eq.~\eqref{eq8}). Note that though the oscillators’ amplitudes decay over time, for a given eigenstate they decay with the same rates. Thus, the ratio ${a_1}/{a_2}$ becomes fixed. As a result, the non-Hermitian system evolves from any initial state to a final state with a fixed ratio of ${a_1}/{a_2}$ determined by the ratio of oscillators’ amplitudes in the eigenstate with the smallest decay rate.

Above the EP, the eigenstates of the non-Hermitian system~\eqref{eq6} have the same decay rates (${\mathop{\rm Re}\nolimits} {\lambda _1} = {\mathop{\rm Re}\nolimits} {\lambda _2}$). In this case, the ratio of ${a_1}/{a_2}$ in a final state depends on the initial state of the system and is determined by the contributions of both eigenstates to the system initial state.

Thus, we conclude that the passing through the EP can be detected by the dependence of ${a_1}\left( {t \to \infty } \right)/{a_2}\left( {t \to \infty } \right)$ on the initial state.

\begin{figure}[t]
  \centering
 \includegraphics[width=\linewidth]{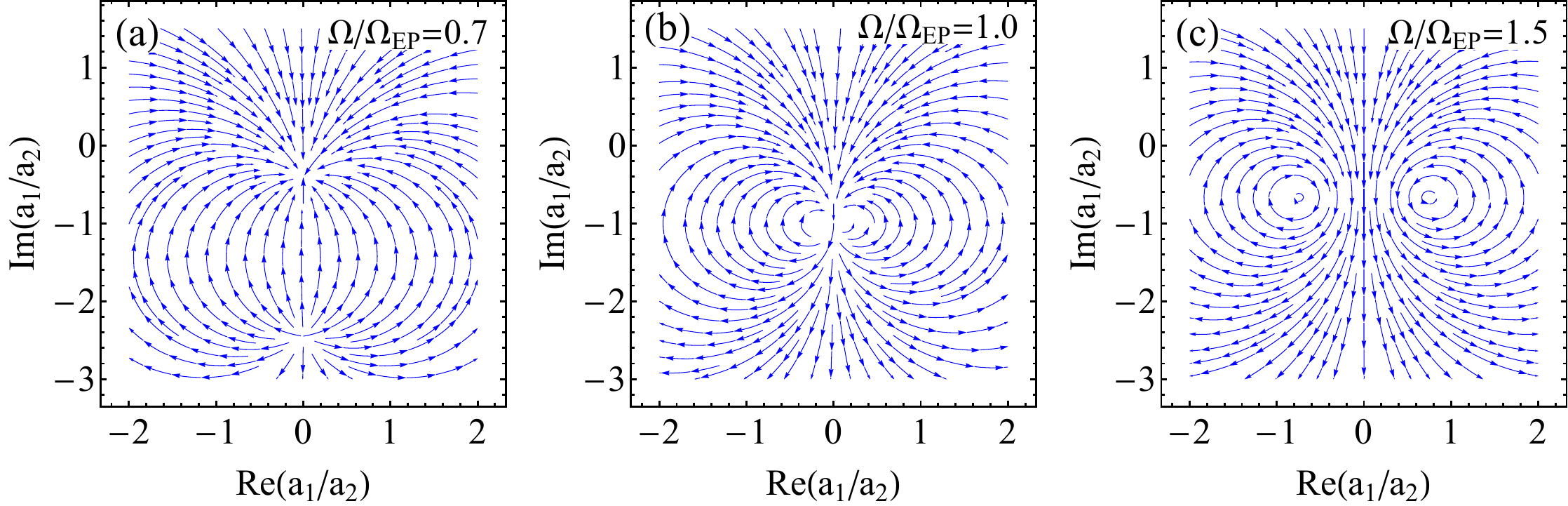}
  \caption{Phase portraits of Eq.~\eqref{eq12} for different values of the coupling strength: $\Omega  = 0.7\,{\Omega _{EP}}$ (a); $\Omega  = {\Omega _{EP}}$ (b); $\Omega  = 1.5\,{\Omega _{EP}}$ (c). Here ${\Omega _{EP}} = \left| {{\gamma _1} - {\gamma _2}} \right|/2 = 0.01{\omega _0}$. Arrows denote the vector $\left( {d{\mathop{\rm Re}\nolimits} A/dt,\,\,d{\mathop{\rm Im}\nolimits} A/dt} \right)$ at the point $\left( {{\mathop{\rm Re}\nolimits} A,{\mathop{\rm Im}\nolimits} A} \right)$.}
  \label{fig:figure3}
\end{figure}

For quantitative description of the EP phase transition, using Eq.~\eqref{eq6} we derive the equation for the quantity $A = {a_1}/{a_2}$:

\begin{equation}
\label{eq11}
\frac{{dA}}{{dt}} = \left( {{\gamma _2} - {\gamma _1}} \right)A + i\Omega {A^2} - i\Omega
\end{equation}
Eq.~\eqref{eq11} can be rewritten as two first-order differential equations for the real and imaginary parts of $A = {a_1}/{a_2}$:

\begin{equation}
\label{eq12}
\begin{array}{l}
\frac{{d{\mathop{\rm Re}\nolimits} A}}{{dt}} = \left( {{\gamma _2} - {\gamma _1}} \right){\mathop{\rm Re}\nolimits} A - 2\Omega {\mathop{\rm Re}\nolimits} A \cdot {\mathop{\rm Im}\nolimits} A\\
\\
\frac{{d{\mathop{\rm Im}\nolimits} A}}{{dt}} = \left( {{\gamma _2} - {\gamma _1}} \right){\mathop{\rm Im}\nolimits} A + \Omega \left( {{{\left( {{\mathop{\rm Re}\nolimits} A} \right)}^2} - {{\left( {{\mathop{\rm Im}\nolimits} A} \right)}^2}} \right) - \Omega 
\end{array}
\end{equation}
The phase portraits of Eq.~\eqref{eq12}, i.e., the trajectories of the system dynamics in the space ${\mathop{\rm Re}\nolimits} A$ and ${\mathop{\rm Im}\nolimits} A$ are shown in Figure~\ref{fig:figure3}.

There are two fixed points of Eq.~\eqref{eq12} in the coordinate plane of ${\mathop{\rm Re}\nolimits} A$ and ${\mathop{\rm Im}\nolimits} A$. These points are defined by following formulas:

\begin{equation}
\label{eq13}
\left( {{\mathop{\rm Re}\nolimits} A,{\mathop{\rm Im}\nolimits} A} \right) = \left( {0,\frac{{{\gamma _1} - {\gamma _2} \pm \sqrt {{{\left( {{\gamma _1} - {\gamma _2}} \right)}^2} - 4{\Omega ^2}} }}{{2\Omega }}} \right),\,\,\,\,\,\,\,\Omega  < {\Omega _{EP}}
\end{equation}

\begin{equation}
\label{eq14}
\left( {{\mathop{\rm Re}\nolimits} A,{\mathop{\rm Im}\nolimits} A} \right) = \left( { \pm \frac{{\sqrt {4{\Omega ^2} - {{\left( {{\gamma _1} - {\gamma _2}} \right)}^2}} }}{{2\Omega }},\frac{{{\gamma _1} - {\gamma _2}}}{{2\Omega }}} \right),\,\,\,\,\,\,\,\,\Omega  \ge {\Omega _{EP}}
\end{equation}
The points correspond to the eigenstates of the non-Hermitian system of Eq.~\eqref{eq6}.

It is seen that below the EP ($\Omega  < {\Omega _{EP}}$), one of the fixed point is an attractive point, while other point is repulsion point (Figure~\ref{fig:figure3}a). As a result, the system evolves from arbitrary initial state to the specified final state. Note that Eq.~\eqref{eq12} and the phase portraits are symmetrical relative to the change ${\mathop{\rm Re}\nolimits} A \to  - {\mathop{\rm Re}\nolimits} A$. Below the EP, the final state is also symmetrical relative to this transformation of variables (in the final state ${\mathop{\rm Re}\nolimits} A = 0$).

Above the EP ($\Omega  > {\Omega _{EP}}$), the system evolves along closed trajectories (Figure~\ref{fig:figure3}c). The ratio of the oscillators’ amplitudes depends on time and the initial state. It is important that, though Eq.~\eqref{eq12} and the phase portraits are symmetrical relative to the change ${\mathop{\rm Re}\nolimits} A \to  - {\mathop{\rm Re}\nolimits} A$, the system trajectories are not. Depending on the initial condition, the system evolution occurs either at positive or at negative values of ${\mathop{\rm Re}\nolimits} A$ (Figure~\ref{fig:figure3}c) and, consequently, the phase difference between the first and second oscillators, $\Delta \varphi  = arg\left( {{a_1}} \right) - arg\left( {{a_2}} \right) = arg\left( {{a_1}/{a_2}} \right)$, lies either in the range $\left( {\pi /2,\,\,3\pi /2} \right)$, or $\left( { - \pi /2,\,\,\pi /2} \right)$.

Thus, the trajectories along which the system evolves lose the symmetry. This behavior is manifestation of a spontaneous symmetry breaking in the non-Hermitian system and corresponds to the EP phase transition.

\subsection{Order parameter for the EP phase transition}
\begin{figure}[t]
  \centering
 \includegraphics[width=0.5\linewidth]{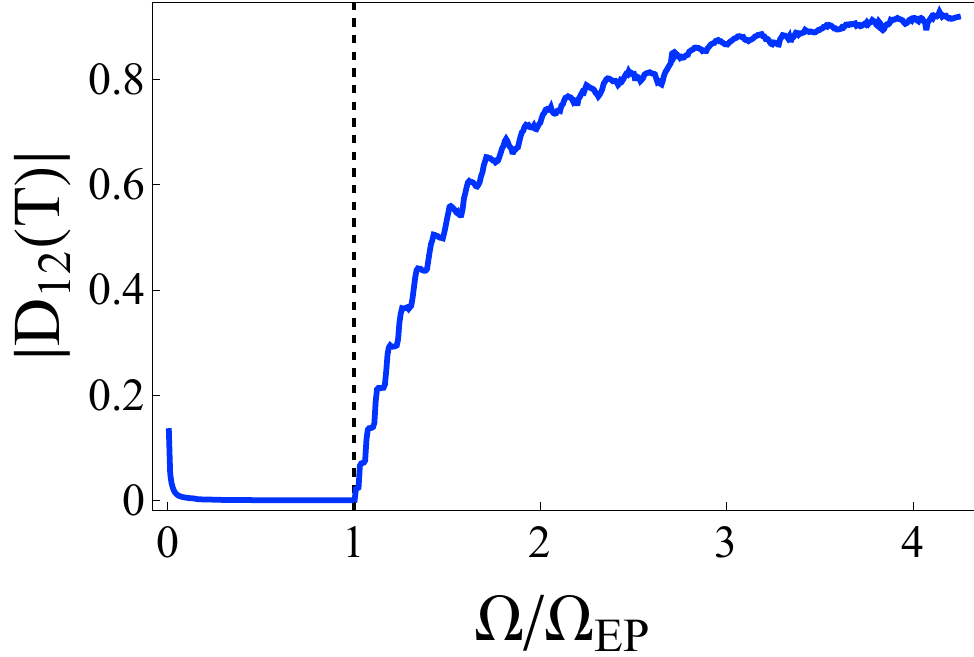}
  \caption{Dependence of the absolute value of the variance ${D_{12}}\left( {T \to \infty } \right)$ on the coupling strength $\Omega$. The vertical dashed line shows the coupling strength corresponding to the EP phase transition.}
  \label{fig:figure4}
\end{figure}

To characterize the EP phase transition, we introduce an order parameter based on the change in the system dynamics. To this end, we calculate the integral ${I_{12}}\left( T \right) = \frac{1}{T}\int_0^T {A\left( t \right)dt} \\ \equiv \frac{1}{T}\int_0^T {{a_1}\left( t \right)/{a_2}\left( t \right)dt}$ for the different initial conditions; $T$ is the observation time. We consider the case $T >  > {T_{R1,R2}}$. Then we find the variance ${D_{12}}\left( T \right) = \left\langle {{I_{12}}{{\left( T \right)}^2}} \right\rangle  - {\left\langle {{I_{12}}\left( T \right)} \right\rangle ^2}$ over the initial states of the system. Below the EP, the system state tends to the eigenstate with the lowest relaxation rate. In this case, the ratio of ${a_1}/{a_2}$ in the final state does not depend on the initial state and, therefore, the value of $\mathop {\lim }\limits_{T \to \infty } {I_{12}}\left( T \right)$ also does not depend on the initial state and $\mathop {\lim }\limits_{T \to \infty }  {{D_{12}}\left( T \right)}  = 0$ (Figure~\ref{fig:figure4}). Above the EP, the eigenstates have the same relaxation rates and the ratio of ${a_1}/{a_2}$ changes over time and the system evolution depends on the initial state (see previous section). In this case, the value of $\mathop {\lim }\limits_{T \to \infty } {I_{12}}\left( T \right)$ depends on the initial state and $\mathop {\lim }\limits_{T \to \infty }  {{D_{12}}\left( T \right)}$ is not zero (Figure~\ref{fig:figure4}).

To calculate the dependence of ${{D_{12}}\left( T \right)}$ on the coupling strength, we numerically solve Eq.~\eqref{eq6} using Runge–Kutta method for different initial states. Hereinafter, we choose the random initial states in the following way. We consider that the initial amplitudes of all reservoirs modes are zero (${b_k}\left( t=0 \right)={c_k}\left( t=0 \right)=0$). We set the initial amplitudes of the oscillators as ${a_1}\left( t=0 \right) = {r_1}\,{\rm{exp}}\left( {2\pi i\,\phi_1 } \right)$ and ${a_2}\left( t=0 \right) = {r_2}\,{\rm{exp}}\left( {2\pi i\,\phi_2 } \right)$, where $r_1$, $r_2$, $\phi_1$ and $\phi_2$ are random real numbers in the interval $\left[ {0,1} \right]$. Then, we average ${I_{12}}\left( T \right)$ and ${I_{12}}\left( T \right)^2$ over the number of the initial states and calculate the variance ${{D_{12}}\left( T \right)}$. The same approach is used further when calculating ${{D_{12}}\left( T \right)}$ in the Hermitian system~\eqref{eq2}-\eqref{eq5}.

Our calculations show that the variance ${{D_{12}}\left( {T \to \infty } \right)}$ experiences qualitative change at the EP (Figure~\ref{fig:figure4}). This change is due to the spontaneous symmetry breaking in the system trajectories occurring at the EP (see Figure~\ref{fig:figure3}). Therefore, the variance ${{D_{12}}\left( {T \to \infty } \right)}$ can serve as an order parameter of the EP phase transition.

\subsection{Signature of EP phase transition in a Hermitian system}
In this section, we study manifestations of the EP phase transitions in Hermitian systems, which include reservoirs of finite sizes. For this purpose, we study the time evolution of the Hermitian system, which is described by Eqns.~\eqref{eq2}–\eqref{eq5}  and consists of two coupled oscillators interacting with their reservoirs. The reservoirs consist of set of ${N_{1,2}}$ oscillators (${N_1} = {N_2} = 40$).

\begin{figure}[t]
  \centering
 \includegraphics[width=\linewidth]{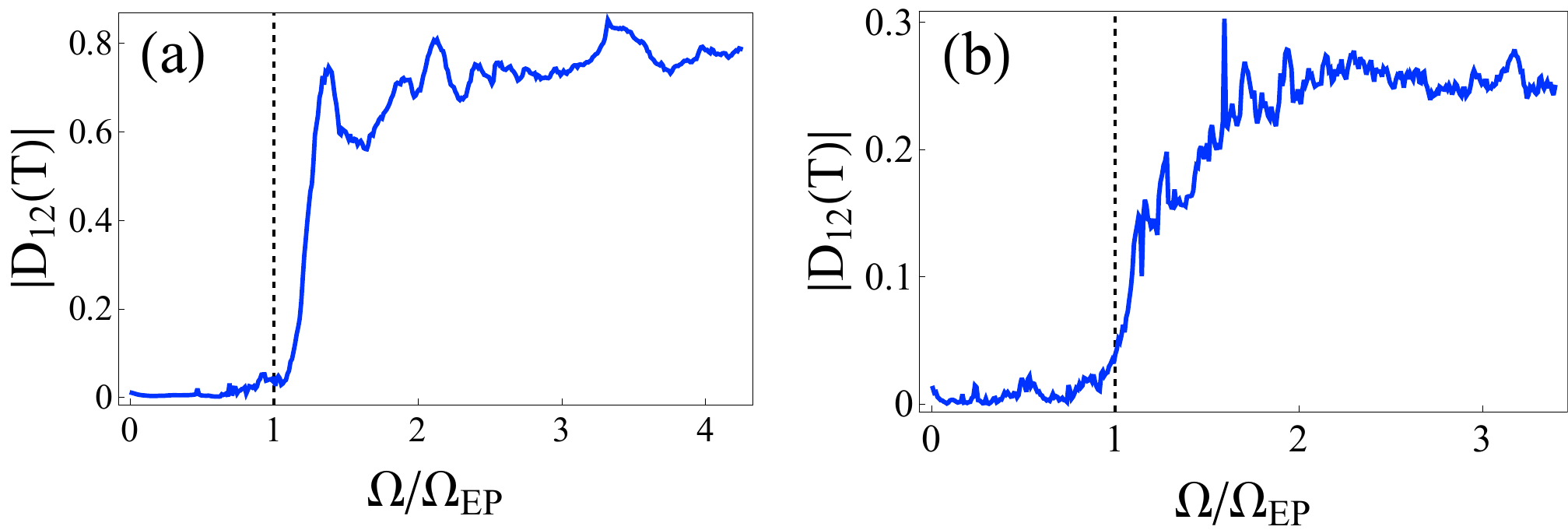}
  \caption{Dependence of the absolute value of the variance ${D_{12}}\left( T \right)$ on the coupling strength in the Hermitian system calculated at time $T = 650\omega _0^{ - 1} \approx 0.5\,{T_{R1,R2}}$ (a) and $T = 13000\omega _0^{ - 1} \approx 10\,{T_{R1,R2}}$ (b). The vertical dashed lines show the coupling strength corresponding to the EP in the non-Hermitian system~\eqref{eq7}, i.e. $\Omega = {\Omega _{EP}} =  {{\left| {{\gamma _1} - {\gamma _2}} \right|}}/{2}$, where ${\gamma _{1,2}} = {{\pi g_{1,2}^2}}/{{\delta {\omega _{1,2}}}}$ ~\eqref{eq10}. The number of initial states, over which the averaging is made, is 800 (a) and 300 (b).}
  \label{fig:figure5}
\end{figure}

Similar to the case of non-Hermitian system, we calculate the variance ${D_{12}}\left( T \right) = \left\langle {{I_{12}}{{\left( T \right)}^2}} \right\rangle  - {\left\langle {{I_{12}}\left( T \right)} \right\rangle ^2}$ over the initial states of the Hermitian system. At time much smaller than the return times ($t <  < {T_{R1,R2}}$) the evolution of Hermitian system approximately coincides with the one of the non-Hermitian system (see Figure~\ref{fig:figure2}). It is not surprising that the variance ${D_{12}}\left( T \right)$ calculated at time $T <  < {T_{R1,R2}}$ demonstrates the same behavior as the one in the non-Hermitian system (cf. Figure~\ref{fig:figure4} and Figure~\ref{fig:figure5}a). Thus, at $t <  < {T_{R1,R2}}$, the EP phase transition manifests itself in the behavior of Hermitian system and the variance continues to play the role of the order parameter.

\begin{figure}[t]
  \centering
 \includegraphics[width=\linewidth]{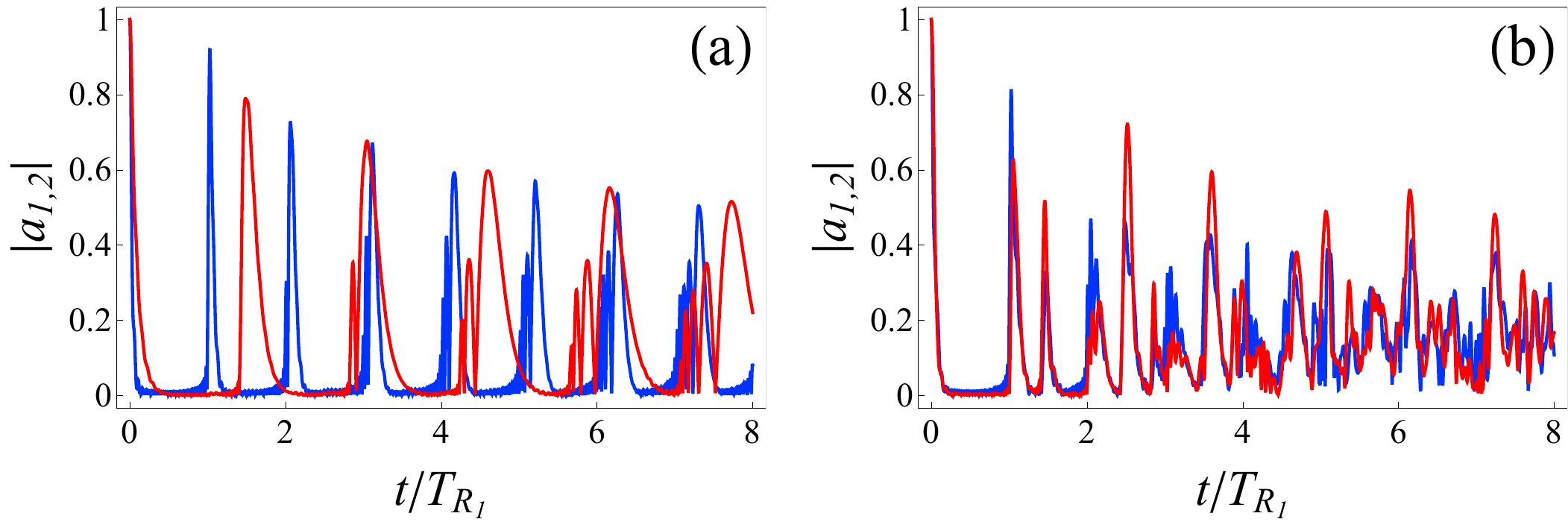}
  \caption{Dependencies of the absolute values of the amplitudes of the first (blue line) and second (red line) oscillators on time calculated by Eqns.~\eqref{eq2}–\eqref{eq5} when the coupling strength $\Omega  = {10^{ - 4}}{\omega _0} \approx 0.01{\Omega _{EP}}$ (a); $\Omega  = 2 \cdot {10^{ - 2}}{\omega _0} \approx 2.12\,{\Omega _{EP}}$ (b). Here ${\Omega _{EP}}=3\pi  \cdot {10^{ - 3}}{\omega _0}$ is defined by analogy with the one in the non-Hermitina system~\eqref{eq7}, i.e., ${\Omega _{EP}} =  {{\left| {{\gamma _1} - {\gamma _2}} \right|}}/{2}$, where ${\gamma _{1,2}} = {{\pi g_{1,2}^2}}/{{\delta {\omega _{1,2}}}}$~\eqref{eq10}. ${T_{R1}} = 2\pi /\delta {\omega _1}$ is the return time of the first reservoir. The following parameters are used ${N_1} = {N_2} = 40$; $\delta {\omega _1} = 5 \times {10^{ - 3}}\,{\omega _0}$; $\delta {\omega _2} = 5 \times {10^{ - 3}}\,{\omega _0}/\sqrt 2$; ${g_1} = 2\sqrt {10}  \times {10^{ - 3}}\,{\omega _0}$; ${g_2} = \sqrt {10}  \times {10^{ - 3}}\,{\omega _0}$. The initial state is ${a_1}\left( 0 \right) = 1$; ${a_2}\left( 0 \right) = 1$; ${b_k}\left( 0 \right) = 0$; ${c_k}\left( 0 \right) = 0$.}
  \label{fig:figure6}
\end{figure}

At time greater than the return times ($t > {T_{R1,R2}}$) the dynamics of Hermitian system qualitatively differ from the one of the non-Hermitian system (Figure~\ref{fig:figure2}). Instead of the exponential decay, the collapses and revivals of the oscillations take place in the system. At greater times ($t >  > {T_{R1,R2}}$), the collapses and revivals are mixed and the system dynamics become complex (Figure~\ref{fig:figure6}).

Since at $t >  > {T_{R1,R2}}$ the behavior of Hermitian system qualitatively differs from the one of the non-Hermitian system, it is difficult to expect that the EP phase transition manifests itself in the dynamics of Hermitian system in this case. However, our calculations show that the signature of EP phase transition is visible even at $t >  > {T_{R1,R2}}$. To demonstrate this, we calculate the variance ${D_{12}}\left( T \right)$ for times much greater than the return times (Figure~\ref{fig:figure5}b). It is seen that when passing through the threshold coupling strength the variance ${D_{12}}\left( {T >  > {T_{R1,R2}}} \right)$ changes similar to the one in the non-Hermitian system (Figure~\ref{fig:figure5}b). The threshold coupling strength equals ${\Omega _{EP}}$, i.e., corresponds to the EP in the non-Hermitian system (Eq.~\eqref{eq7}), where the relaxation rates are determined by Eq.~\eqref{eq10}.

Thus, the variance ${D_{12}}\left( {T >  > {T_{R1,R2}}} \right)$ sharply increases at the coupling strength corresponding to the EP in the non-Hermitian system. This behavior indicates that there is a transition manifesting in the dynamics of Hermitian system that is similar to the EP phase transition in the non-Hermitian system. The change occurring at such a transition is visible in the dynamics of Hermitian system, in which the reservoirs have different frequency steps ($\delta {\omega _1} \ne \delta {\omega _2}$). In this system, the return times of the first and second reservoirs are different from each other. When the coupling strength between the oscillators is smaller than ${\Omega _{EP}}$, the revivals of oscillations in the first and second oscillators occur at different times (Figure~\ref{fig:figure6}a). When the coupling strength between the oscillators is greater than ${\Omega _{EP}}$, the dynamics of oscillators become similar to each other (Figure~\ref{fig:figure6}b).

Thus, we conclude that the EP phase transition manifests itself in the dynamics of Hermitian system at times both smaller and greater than the return times. Note that in the Hermitian system, the interaction of the oscillators with finite number of the reservoirs’ oscillators introduces randomness into the system dynamics (Figure~\ref{fig:figure6}). At the same time, the increase in the coupling strength, $\Omega$, makes the oscillators' dynamics more matched (Figure~\ref{fig:figure6}b). When $\Omega>\Omega_{EP}$, a partial matching of oscillators, similar to the one in the non-Hermitian system (see Figure~\ref{fig:figure3}c), takes place in the Hermitian system. We interpret this behavior as a signature of the exceptional point phase transition. This transition has common features with the ones in other physical systems, in which the competition between the coupling of the subsystems and its interaction with the environment can lead to phase transitions. For example, in a lattice of spins, the increase in the coupling strength between them can compensate for thermal fluctuations and lead to consistency in the spins' orientations resulting in the phase transition \cite{ref54L}.

\subsection{Influence of number of reservoirs degrees of freedom on the manifestations of the EP phase transition}

In the previous section, we demonstrated that there is a threshold in the dynamics of Hermitian system, which can be associated with the EP phase transition in the non-Hermitian system. This transition manifests itself even at time much greater than the return time. The return times $T_{R1,R2}$ are inversely proportional to the differences of frequencies between the neighboring modes in the reservoirs, $\delta {\omega _{1,2}}$. In this section, we study the influence of the density of the modes ($\sim 1/\delta {\omega _{1,2}}$) on the behavior of the order parameter. We fix the frequency range in which the reservoirs modes lie (i.e., we fix products $N_{1,2} \delta {\omega _{1,2}}$). In this case, the return times are proportional to the number of reservoirs modes. For this reason, to compare the behavior of systems with different numbers of the reservoirs modes, it is necessary to change the observation time along with the change in the number of the modes. In addition, a change of $\delta {\omega _{1,2}}$ ($N_{1,2}$) leads to an alteration of the exponential decay rate $\gamma_{1,2}=\pi g_{1,2}^2/\delta{\omega_{1,2}}$~\eqref{eq10}. To compare systems with a different numbers of the reservoirs modes, we change $g_{1,2}$ along with the change in $\delta_{\omega_{1,2}}$, so that $\gamma_{1,2}$ are the same for all systems.

\begin{figure}[t]
  \centering
 \includegraphics[width=0.5\linewidth]{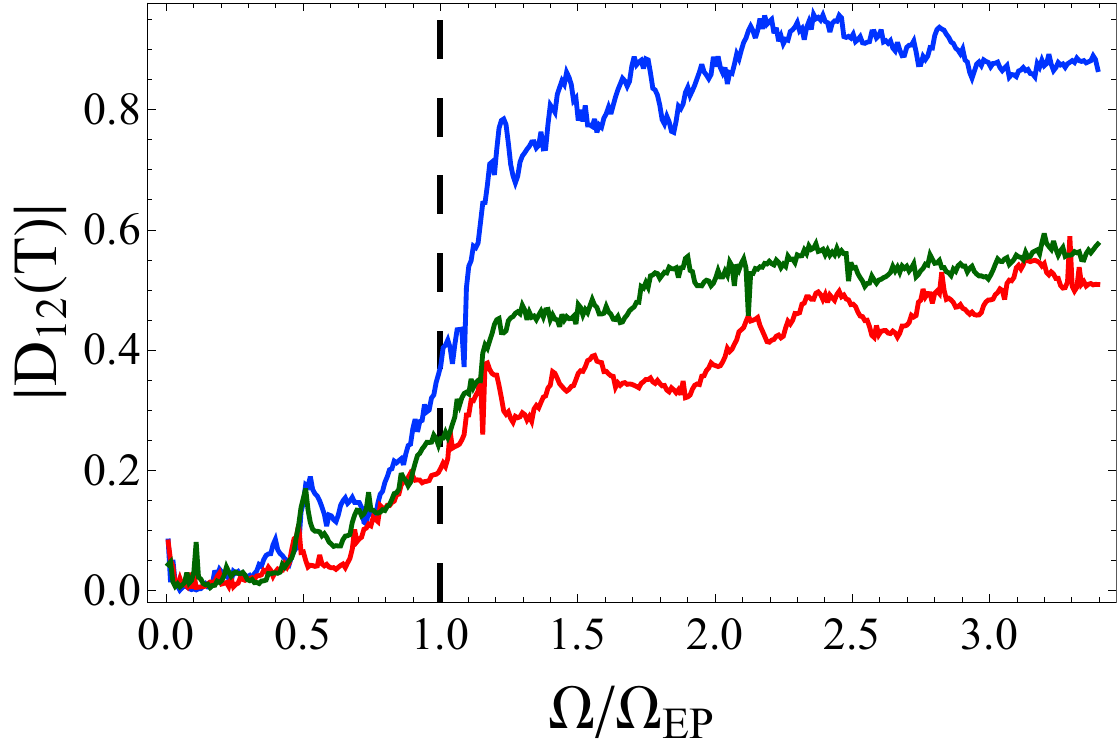}
  \caption{Dependence of the absolute value of the variance ${D_{12}}\left( T \right)$ on the coupling strength in the Hermitian system calculated at time $T = 13000\omega _0^{ - 1} \approx 10\,{T_{R1,R2}}$. The numbers of the reservoirs modes are $N_1 = N_2 = 40$ (the blue line); $N_1 = N_2 = 30$ (the green line); $N_1 = N_2 = 20$ (the red line). The vertical dashed lines show the coupling strength $\Omega = {\Omega _{EP}} =  {{\left| {{\gamma _1} - {\gamma _2}} \right|}}/{2}$~\eqref{eq7}, where ${\gamma _{1,2}} = {{\pi g_{1,2}^2}}/{{\delta {\omega _{1,2}}}}$ ~\eqref{eq10}. The number of initial states, over which the averaging is made, is 100.}
  \label{fig:figure6and3/4}
\end{figure}

\begin{figure}[t]
  \centering
 \includegraphics[width=\linewidth]{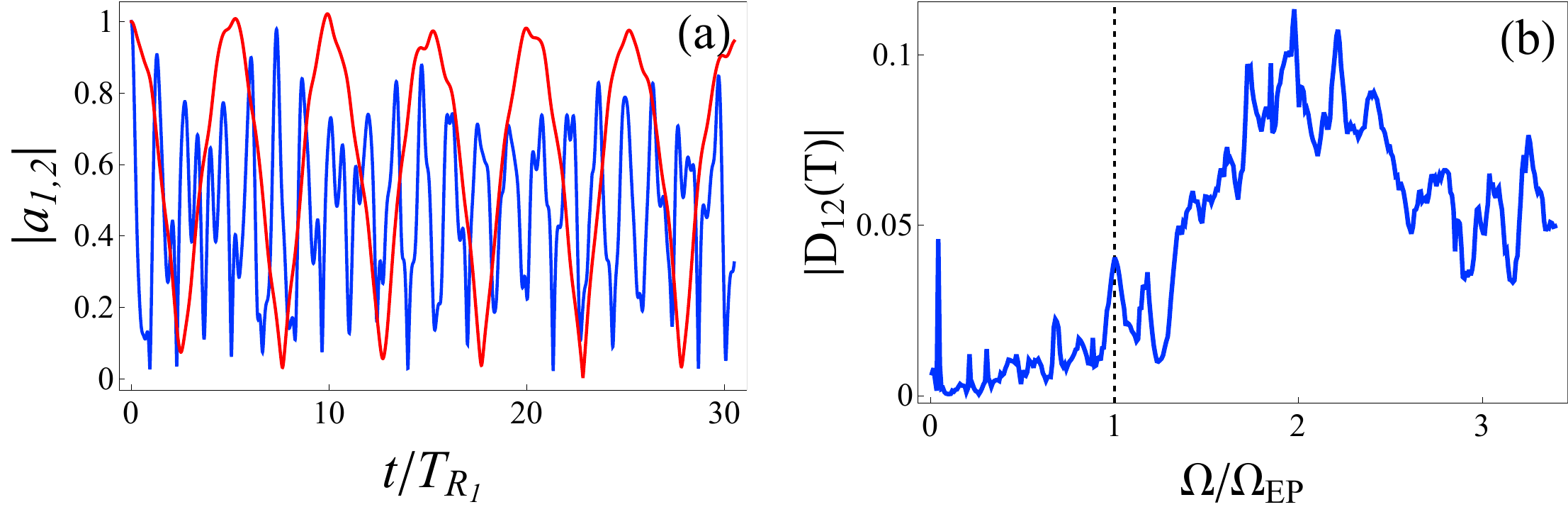}
  \caption{(a) Dependencies of the absolute values of the amplitudes of the first (blue line) and second (red line) oscillators on time calculated by Eqns.~\eqref{eq2}–\eqref{eq5}. The oscillators interact with the reservoirs of ${N_1} = {N_2} = 4$ oscillators. ${T_R} = 2\pi /\delta \omega$ is the return time of the reservoirs. $\delta {\omega _{1,2}} = \delta \omega  = 5 \cdot {10^{ - 2}}\,{\omega _0}$; ${g_1} = 2.5 \cdot {10^{ - 2}}{\omega _0}$;  ${g_2} = 1.5 \cdot {10^{ - 2}}{\omega _0}$; $\Omega  = {10^{ - 3}}{\omega _0}$. (b) Dependence of the absolute value of the variance ${D_{12}}\left( T \right)$ on the coupling strenght  $\Omega$. Here $T = 30\,{T_R}$; the number of initial states, over which the averaging is made, is 500. ${N_1} = {N_2} = 4$; ${g_1} = 2.5 \cdot {10^{ - 2}}{\omega _0}$; ${g_2} = 1.5 \cdot {10^{ - 2}}{\omega _0}$; $\delta {\omega _{1,2}} = \delta \omega  = 5 \cdot {10^{ - 2}}\,{\omega _0}$. The black dashed line shows the coupling strength corresponding to the EP in the non-Hermitian system~\eqref{eq7}, i.e. $\Omega = {\Omega _{EP}} =  {{\left| {{\gamma _1} - {\gamma _2}} \right|}}/{2}$, where ${\gamma _{1,2}} = {{\pi g_{1,2}^2}}/{{\delta {\omega _{1,2}}}}$ ~\eqref{eq10}.}
  \label{fig:figure7}
\end{figure}

Our calculations based on the Equations~\eqref{eq2}–\eqref{eq5} show that the decrease of the number of reservoirs modes leads to the fact that the transition becomes less clear (see Figure~\ref{fig:figure6and3/4}). As the number of the reservoirs modes decreases further, the return times become shorter than the exponential decay times ($T_{1,2}<<\gamma_{1,2}^{-1}$). In this case, the dynamics of the Hermitian system differ qualitatively from the one of the non-Hermitian system at all times (cf. Figure~\ref{fig:figure2} and Figure~\ref{fig:figure7}a). It is complex and resembles random fluctuations (Figure~\ref{fig:figure7}a). However, even in this regime, the system dynamics demonstrate the transition (Figure~\ref{fig:figure7}a), which occurs when the coupling strength between the oscillators $\Omega \approx {\Omega _{EP}} =  {{\left| {{\gamma _1} - {\gamma _2}} \right|}}/{2}$, where $\gamma_{1,2} = \pi g_{1,2}^2/\delta{\omega_{1,2}}$~\eqref{eq10}. The fact that the transition takes place even when the system dynamics are so complex allows us to argue about it occurs due to the entire temporal dynamics of the system, and not just due to exponential decay stages.

\section{Experimental setup for observation of the signature of EP phase transition in a Hermitian system}

In the previous sections, we demonstrated that there is the signature of EP phase transition in the dynamics of the Hermitian systems consisting of two coupled oscillators and the reservoirs with finite number of the modes. Here we discuss possible experimental schemes for observing this transition in systems with several tens of degrees of freedom. Such systems should consist of two coupled bosonic subsystems, at least one of which interacts with a multimode system simulating the reservoir with finite number of degrees of freedom. High-Q optical cavities \cite{ref54}-\cite{ref58}, superconducting qubits \cite{ref59}-\cite{ref66} can play the role of such bosonic subsystems. In turn, the finite length waveguides or multimode cavities can serve as reservoirs with a finite number of degrees of freedom. The length of waveguide or cavity determines the step between the frequencies in such a reservoir.

\begin{figure}[t]
  \centering
 \includegraphics[width=\linewidth]{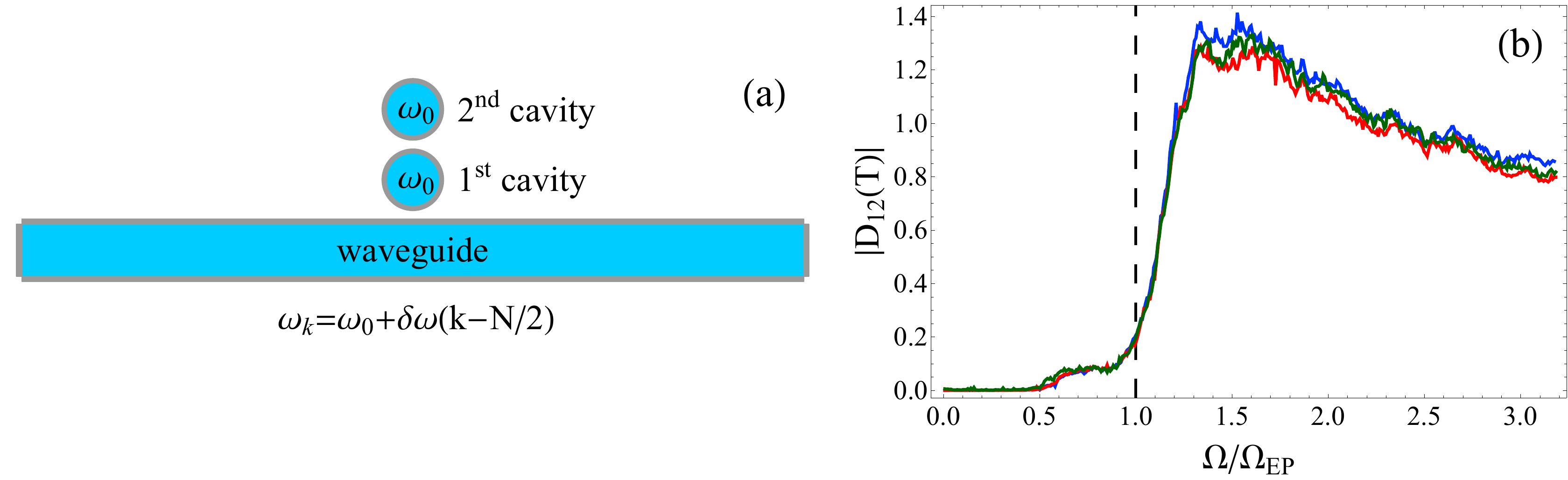}
  \caption{(a) Scheme of a photonic system for observing the signature of EP phase transition. (b) The dependence of the absolute value of the variance ${D_{12}}\left( T \right)$ on the coupling strength between the optical cavities $\Omega$ calculated by Eqns.~\eqref{eq15}–\eqref{eq18}. The blue line $\Gamma T_{R}=0$; the green line $\Gamma T_{R}= 0.1$; the red line $\Gamma T_{R}= 1$. $T_{R}=2 \pi/\delta \omega$; $N = 40$; $\delta \omega  = 5 \cdot {10^{ - 3}}{\omega _0}$; $g = 2\sqrt {10}  \times {10^{ - 3}}{\omega _0}$. The black dashed line shows the coupling strength $\Omega = {\Omega _{EP}} =  {{\left| {{\gamma _1} - {\gamma _2}} \right|}}/{2}$~\eqref{eq7}, where ${\gamma _{1,2}} = {{\pi g_{1,2}^2}}/{{\delta {\omega _{1,2}}}}$ ~\eqref{eq10}.}
  \label{fig:figure8}
\end{figure}

As an example of system suitable for observing the signature of EP phase transition in the non-Markovian regime, we propose optical circuit consisting of two coupled optical cavities, one of which interacts with a finite length waveguide (Figure~\ref{fig:figure8}a). In such a system, the interaction of cavity field with the finite length waveguide field leads to the collapses and the revivals of the energy in the cavities. In this system, the return process is due to the reflection of the electromagnetic waves from the waveguide' ends. Indeed, in the waveguide with the linear frequency dispersion, the frequencies of the standing modes are determined by the expression ${\omega _k} = \frac{{\pi c}}{L}k = \delta \omega \,k$, where $\delta \omega  = \frac{{\pi c}}{L}$ and $k$ is an integer number. The interaction of the cavity with the waveguide' modes leads to excitation of the electromagnetic waves, which propagate along the waveguide, reflect from the waveguide ends and return to the cavity. The return time is given by the expression ${T_R} = 2L/c = 2\pi /\delta \omega $, which coincides with the one obtained by the consideration of the phase condition.

To illustrate the signature of EP phase transition in this system, based on Eqns.~\eqref{eq2}–\eqref{eq5}, we calculate the time dependence of light amplitudes in the cavities. Our calculations show that the variance ${D_{12}}\left( T \right)$ demonstrates the threshold dependence on the coupling strength between the cavities (Figure~\ref{fig:figure8}b). Thus, the variance ${D_{12}}\left( T \right)$ plays the role of the order parameter for the transition in such a system.

The losses are always present in real physical systems. To determine the influence of the losses on the transition in the considered system, we added the relaxation terms in the Equations for the Hermitian system~\eqref{eq2}–\eqref{eq5}  
\begin{equation}
\label{eq15}
\frac{{d{a_1}}}{{dt}} =  - i{\omega _0}{a_1} - \Gamma {a_1}- i\,\Omega {a_2} - i\sum\limits_{k = 1}^{{N_1}} {{g_1}{b_k}}
\end{equation}

\begin{equation}
\label{eq16}
\frac{{d{a_2}}}{{dt}} =  - i{\omega _0}{a_2} - \Gamma {a_2}- i\,\Omega {a_1} - i\sum\limits_{k = 1}^{{N_2}} {{g_2}{c_k}}
\end{equation}

\begin{equation}
\label{eq17}
\frac{{d{b_k}}}{{dt}} =  - i\omega _k^{\left( 1 \right)}{b_k} -\Gamma {b_k}- i\,{g_1}{a_1}
\end{equation}

\begin{equation}
\label{eq18}
\frac{{d{c_k}}}{{dt}} =  - i\omega _k^{\left( 2 \right)}{c_k}\, -\Gamma {c_k} - i\,{g_2}{a_2}
\end{equation}
Here $\Gamma$ is the relaxation rate of the cavity and waveguide's modes (for simplicity we assume that they are equal).

We consider the two cases: $\Gamma T_{R}<<1$ and $\Gamma T_{R}=1$ (Figure~\ref{fig:figure8}b). In the first case, the system dynamics demonstrate a large number of collapses and revivals ($\sim 1/(\Gamma T_{R})$). While, in the second case, only the exponential decay takes place. However, our calculations show that in the both cases the variance ${D_{12}}\left( T \right)$ demonstrates the threshold dependence on the coupling strength. This indicates the robustness of the predicted transition to losses.

Note that besides the proposed optical scheme, there are a number of other circuits that are suitable for observation of the signature of EP phase transition. For example, in the work \cite{ref66}, the authors investigated the system of Xmon qubits strongly coupled to a slow-light phonon waveguide. They demonstrated revivals and collapses in the system dynamics \cite{ref66}. Such systems based on superconducting qubits are also suitable for observing the signature of EP phase transition in the non-Markovian regime.

Thus, we conclude that the signature of EP phase transition in the non-Markovian regime can be observed in the different types of systems including photonics and qubit systems.

\section{Conclusions}
To conclude, we consider the Hermitian system consisting of the two coupled oscillators and the reservoirs interacting with the oscillators. Usually, the non-Hermitian equations are used to describe the behavior of this system. The non-Hermitian equations are obtained by the elimination of the degrees of freedom of the reservoirs, which are usually considered with infinite number of degrees of freedom \cite{ref49,ref50}. The resultant non-Hermitian system has an exceptional point (EP), at which two eigenstates coalesce and their eigenvalues coincide. The passing through the EP leads to the spontaneous symmetry breaking of eigenstates, which is associated with the EP phase transition. In the case of reservoirs with a finite number of degrees of freedom, the reverse energy flow from the environment into the system appears that leads to collapses and revivals of the amplitude oscillations in the system. As a result, the dynamics of Hermitian system, which includes both the oscillators and the reservoirs, cannot be described by the non-Hermitian equations. The manifestation of the EP phase transitions in such Hermitian systems early is not discussed.

To solve this problem, we study the behavior of the Hermitian system of two coupled oscillators interacting with the reservoirs with only few tens of degrees of freedom. We show that the dynamics of this Hermitian system demonstrate a transition occurring at the coupling strength between the oscillators corresponding to the EP in the non-Hermitian system. This transition is manifested in the change in the system dynamics, which observes even at time much greater than the return time when the dynamics of Hermitian system differ qualitatively from the one of non-Hermitian system. We find an order parameter characterizing the transition in both the non-Hermitian and Hermitian system. Our results demonstrate the existence of a new class of the transitions that occur in Hermitian systems and establish a connection between it and the EP phase transitions in non-Hermitian systems. We propose the photonic circuit for observing such a transition.

\section*{Acknowledgments}
The study was financially supported by a grant from Russian Science Foundation (project No. 20-72-10057). T.T.S. and Yu.E.L. thanks foundation for the advancement of theoretical physics and mathematics “Basis”. Yu.E.L. acknowledged Basic Research Program at the National Research University HSE. 

\bibliographystyle{plain}

\end{document}